\documentclass[10pt,osajnl2,letterpaper,twocolumn,floatfix]{revtex4} 
\usepackage{amsmath}
\usepackage{hyperref}
\usepackage{color}
\usepackage[mathenv]{}
\usepackage{graphicx}
\usepackage{natbib}
\begin{document}

\title{Low-frequency vacuum squeezing via polarization self-rotation in Rb vapor}

\author{Eugeniy E. Mikhailov}
\author{Irina Novikova}
\affiliation{The College of William $\&$ Mary, Williamsburg, Virginia, 23187}

\begin{abstract}

We observed squeezed vacuum light at 795~nm in $^{87}$Rb vapor via resonant
polarization self-rotation, and report noise sidebands suppression of $\approx
1$~dB below shot noise level spanning from acoustic (30~kHz) to MHz
frequencies. This is the first demonstration of sub-MHz quadrature vacuum
squeezing in atomic systems. The spectral range of observed squeezing matches
well typical bandwidths of electromagnetically induced transparency (EIT)
resonances, making this simple technique for generation of optical fields with
non-classical statistics at atomic transitions wavelengths attractive for
EIT-based quantum information protocols applications.
\end{abstract}

\date{\today}
\maketitle


The reliable   and  efficient   generation   of an  electromagnetic  field with
non-classical  statistics  (i.e.  ``squeezed''   light,  or  `` squeezed''
vacuum) is  important for a  number of applications from precision metrology to
quantum information.  Many recently proposed  protocols for controlling and
manipulating  quantum  states of  light  rely on the resonant coherent
interaction of light with atomic ensembles~\cite{lukin03rmp,julsgaard04}. These
applications require sources of light with controllable quantum mechanical
properties in a characteristic bandwidth that is resonant with atomic
transitions. Electromagnetically induced transparency (EIT) resonances, for
example, are widely used in slow light and quantum memory
experiments~\cite{chaneliere_storage_2005, eisaman_electromagnetically_2005},
few photon wave-packet generation and
control~\cite{eisaman04,matsukevich06,chen06,kolchin_generation_2006}, etc.
Their typical bandwidths range from a few tens of Hz to a few
MHz~\cite{lukin03rmp}.

At present optical parametric  oscillators (OPO) operating below threshold
offer the best performance for squeezed vacuum
generation~\cite{bachor_guide_2004}.
Many EIT experiments are based on Rb D${}_1$ line (795nm) transitions, and OPO
performance is limited at this wavelength due to increased material absorption
for $397$~nm near-ultraviolet up-converted pump field, as well as various
photothermal effects arising at higher pump
power~\cite{hetet_squeezed_at_D1_Rb_2007}).
%
Another challenge is  to produce low-frequency squeezing  to match the
characteristic bandwidth of EIT resonances. While the generation of squeezed
light with sub-MHz sideband frequencies is theoretically
possible~\cite{chelkowski_10Hz_squeezing_2007}, practically it becomes a very
challenging and resource-consuming task due to significant experimental
complexity.
Nevertheless, impressive progress in generation of low-frequency squeezed
vacuum at Rb resonance wavelength has been recently
reported~\cite{hetet_squeezed_at_D1_Rb_2007}, and such sources were used to
demonstrate slow light and reversible mapping of squeezed vacuum states via EIT
in Rb~\cite{akamatsu_ultraslow_2007, appel_sq_quantum_memory_Rb_2007,
honda_sq_storage_in_Rb_2007}.

\begin{figure}[h]
  \includegraphics[angle=0, width=1.00\columnwidth]{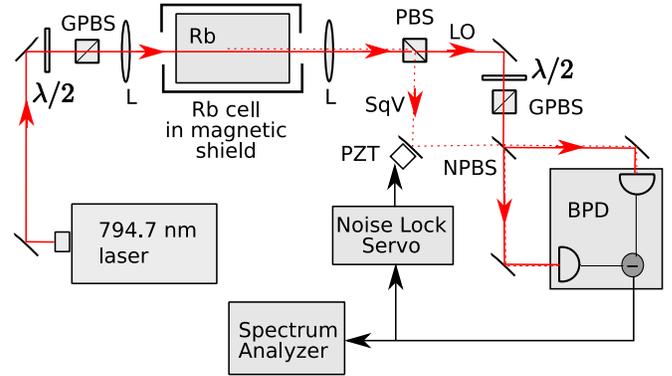}
  \caption{
    Schematic of the experimental setup. SqV is squeezed vacuum field, LO is local
    oscillator. Please see the text for other abbreviations.
    }
  \label{fig:apparatus}
\end{figure}

The generation  of  resonant squeezed vacuum  based  on the nonlinear
interaction of light  with  atoms  offers a simpler  alternative  to
traditional nonlinear crystal-based   squeezers, and various techniques have
been explored recently (see review in~\cite{lezama_numerical_2008}). This
Letter reports successful observation  of low-frequency squeezed vacuum at the
Rb optical D${}_1$ transition ($795~$nm) based on nonlinear polarization
self-rotation effect, recently proposed by Matsko  {\it et
al.}~\cite{matsko_vacuum_2002}. Ries  {\it et al.} reported the
proof-of-principle demonstration on Rb D${}_2$
line~\cite{ries_experimental_2003}, although some later experiments failed to
reproduce this result~\cite{hsu_effect_2006}. The present experiment is aimed
to resolve this controversy.

The polarization   self-rotation  effect   describes   the   rotation  of
polarization direction of elliptically polarized light as it propagates through
a medium, and it occurs  in  many  optical  substances. The effect is
characterized by a self-rotation parameter $g$, such that
$\phi_{SR}=g\varepsilon L$, where $\phi_{SR}$ is the polarization rotation
angle of the input field with ellipticity $\varepsilon$ after traversing
optical medium of length $L$. In resonant atomic vapor self-rotation occurs due
to unbalanced ac-Stark shifts caused by unequal intensities of circularly
polarized components  of the input light field~\cite{novikova_ac-stark_2000,
rochester_self-rotation_2001}.
%
In case of linearly polarized pump field there is no macroscopic polarization
rotation, but same mechanism couples quantum noise in two initially independent
circular components, and thus produces cross-phase modulation between classical
linearly polarized pump field and vacuum field in the orthogonal polarization.
As a result, quadrature squeezing of the vacuum field
occurs~\cite{matsko_vacuum_2002}. The expected noise suppression below standard
quantum limit is proportional to $1/(gL)^2$, but  is reduced by optical losses
in the system. In practice, the observation of maximum squeezing requires the
optimization of many experimental parameters such as laser detuning and power,
atomic density, etc. Spontaneous emission noise in thermal vapor may also
reduce or destroy squeezing by introducing extra noise~\cite{hsu_effect_2006,
lezama_numerical_2008}. We also observed that squeezing generation is sensitive
to the presence of uncompensated magnetic field inside the cell.
%
%
%
%

The schematic of the experiment is shown in Fig.~\ref{fig:apparatus}. The
experimental settings  are  very simple with no need for expensive equipment,
such as powerful  lasers or high-quality optical cavity, and the resulting
non-classical field is automatically generated at near-resonant wavelength.  An
external cavity diode laser ($\approx 7$~mW total power) was tuned to the Rb
$D_1$ line $5^2S_{1/2}\rightarrow5^2P_{1/2}$ ($\lambda \simeq 795$~nm). The
laser beam was focused with a pair of lenses (L) inside a cylindrical  glass
cell, containing isotropically enriched $^{87}$Rb and 2.5~Torr of Ne buffer
gas, and the estimated minimum laser beam diameter inside the cell was
$0.175\pm0.015$~mm FWHM. The cell length and diameter were 75~mm and 22~mm
consecutively, and the cell windows were tilted at about 10$^\circ$ to prevent
backward reflections.
The cell was mounted inside a three-layer magnetic shielding to minimize stray
magnetic fields. The cell was maintained at $66^\circ$C.
Before entering the cell the laser beam passed through a high quality Glan
polarizing beam splitter (GPBS) to purify its linear polarization. A half-wave
plate ($\lambda/2$) placed before the input polarization beam splitter allowed
for smooth adjustments of the pump field intensity.

After the cell the electromagnetic field in orthogonal polarization (squeezed
vacuum field, SqV) was separated on a polarization beam splitter (PBS), and its
noise properties were analyzed using a homodyne detection. The original pump
field played the role of a local oscillator (LO), that was attenuated and
brought to the same polarization as the vacuum field using  another GPBS and a
half-wave plate combo. Phase difference between the local oscillator and the
vacuum field was controlled by a mirror placed or a piezo-ceramic transducer
(PZT) that allowed changing the differential pathlength between two arms.  We
then mixed these two fields at a 50/50 non-polarizing beam splitter(NPBS), and
directed two beams to a home-made balance photodetector (BPD) with  a gain of
$10^4$~V/A, 1~MHz 3~dB bandwidth and electronic noise floor located at 6~dB
below shot noise at low frequencies. Two matched Hamamatsu S5106 photodiodes
with quantum efficiency $\eta=87$\%  and a low noise hight bandwidth TI OPA842
operational amplifier were crucial components of the BPD. For our measurement
we locked the relative phase in BPD detection scheme  to minimum value of
quadrature noise in SqV channel by using a noise-locking
technique~\cite{mckenzie_quantum_2005}. The output of BPD was bandpass filtered
with a  central frequency 1.2~MHz and a RBW 100~kHz and then sent to an
envelope detector with VBW=30Hz,  while dithering the LO phase with a 1~kHz
modulation  frequency. A  demodulation of the output of the envelope detector
with a lock-in amplifier provided correction signal for a servo.


\begin{figure}[h]
  \includegraphics[angle=0,
    width=1.0\columnwidth]{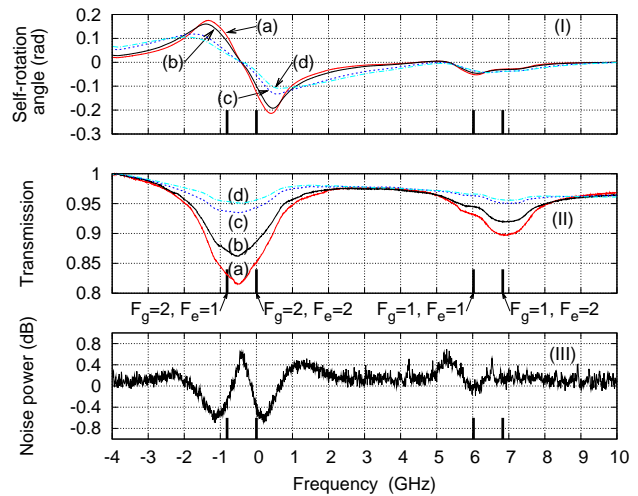}
  \caption{
    Polarization self-rotation angle ($\phi_{SR}$) (I) and transmission (II) in
    the $^{87}$Rb cell  for the light field with $\varepsilon=4^\circ$ ellipticity.
    Laser powers are
    1.04~mW (a), 1.54~mW (b), 4.15~mW (c), and 6.37~mW (d). (III) Minimum
    quadrature noise in orthogonal polarization at 1200~MHz central frequency, 
        for linearly polarized pump field with power 6.58~mW. Zero laser detuning corresponds to $F_g=2 \to F_e=2$
    $^{87}$Rb transition.  }
  \label{fig:squeezing_vs_detuning}
\end{figure}

Fig.\ref{fig:squeezing_vs_detuning}~(I,II) shows the self-rotation and
absorption of the pump field when a small ellipticity $\varepsilon=4^\circ$
 was introduced by a quarter-wave plate placed before
the cell (we define $\varepsilon$ as an angle between $\lambda/4$ plate fast
axis and pump field polarization direction). Maximum self-rotation was observe
near both $F_g=2\rightarrow F_e=1,2$ transitions of ${}^{87}$Rb, although the
polarization ellipse rotated in opposite directions. Increasing the laser power
resulted in increased transparency of the atomic vapor due to more effective
optical pumping of atoms into non-interacting combination of Zeeman sublevels
of $F_g=2$ state as well as to off-resonant $F_g=1$ state. The second mechanism
was most likely also responsible for some reduction in the self-rotation.

Fig.\ref{fig:squeezing_vs_detuning}~(III) shows a sample spectrum of minimum
(squeezed) quadrature of the optical field of orthogonal polarization in case
of linearly polarized pump field (\emph{i.e.} no measurable polarization
self-rotation occurs). The positions of squeezing peaks correspond roughly to
those of self-rotation. A maximum squeezing of 0.6~dB was detected at detunings
of about 100~MHz to the red from $F_g=2 \to F_e=1$ as well as about 100~MHz to
the blue from $F_g=2 \to F_e=2$, and excess noise was observed in the region
between two transitions. We also observed minute amount of squeezing near
$F_g=1 \to F_e=1$ transition.
Vacuum squeezing showed similar dependence of the optical frequency for
different pump powers.

\begin{figure}[h]
\includegraphics[angle=0,width=1.0\columnwidth]{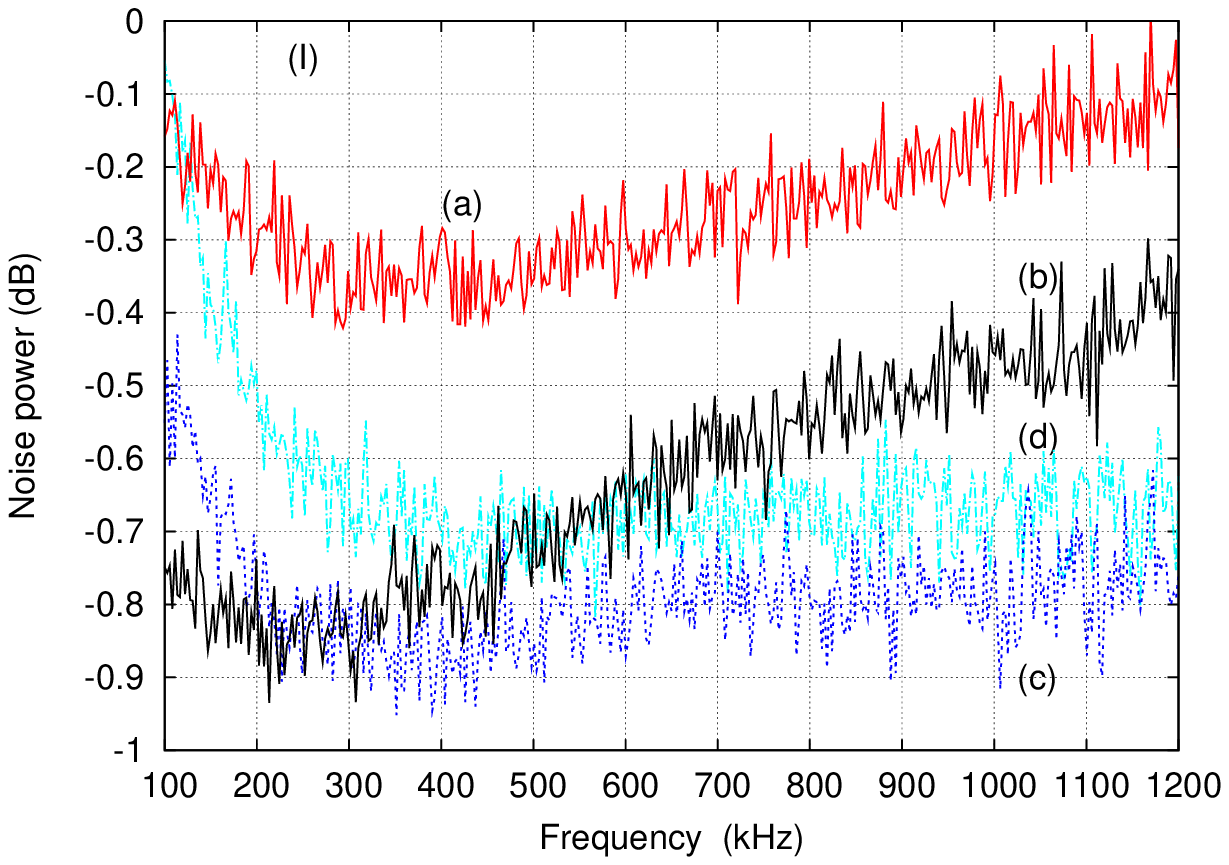}
\includegraphics[angle=0,width=1.0\columnwidth]{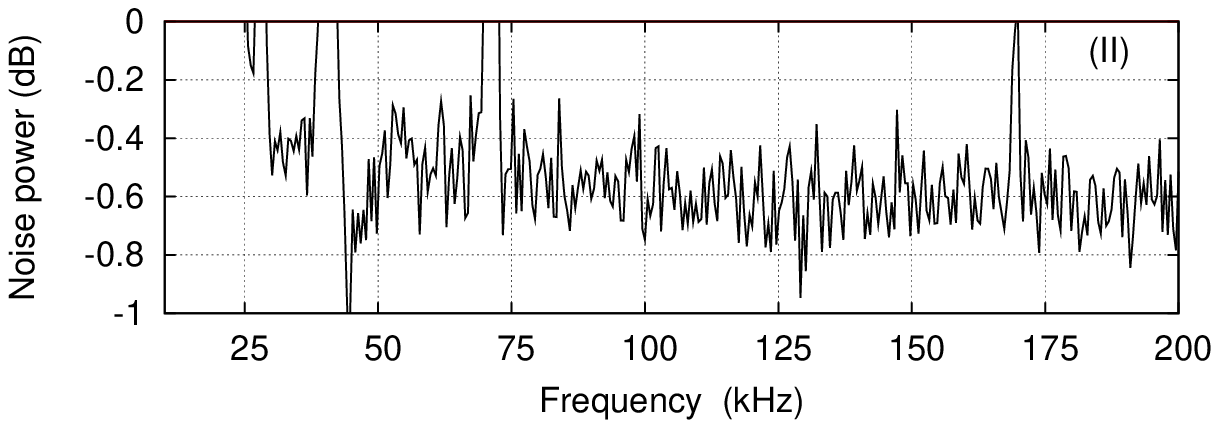}
\caption{
    (I) Squeezed quadrature noise vs sideband frequency for different laser power
    settings. Laser powers are
    1.04~mW (a), 1.54~mW (b), 4.15~mW (c), and  6.58~mW (d).
    (II) Low-frequency component of squeezed quadrature noise spectrum for the laser power 1.54~mW.
     Shot noise corresponds to 0~dB. The laser is tuned to the
    maximum squeezing near 100~MHz detuning.
} \label{fig:squeezing_vs_sideband_frequency}
\end{figure}

At  the   same  time   the  squeezed   quantum  noise   frequency  spectrum
showed   rather    strong   variation   with   the    pump   field   power,
as   Fig.~\ref{fig:squeezing_vs_sideband_frequency}(I)  demonstrates.   The
detected  squeezing was  uniformly  low  at low  laser  powers $>1$~mW.  As
the  pump power  increased,  we  first observed  maximum  squeezing in  the
low-frequency part of the quantum  noise spectrum. Then broadband squeezing
kept increasing at the expense  of low-frequency components and reached its
maximum  at  about 4~mW.  Further  power  increase began  slowly  degrading
observed  squeezing.  Such a  low  value  of  optimal  laser power  may  in
principle  explain why  squeezing was  not observed  in the  experiments of
Hsu  {\it  et al.},  where  much  higher  laser  powers (35~mW)  were  used
~\cite{hsu_effect_2006}.

To investigate the low-frequency part of the squeezing spectrum in more detail,
we studied the case of 1.54~mW pump power more carefully.
Fig.~\ref{fig:squeezing_vs_sideband_frequency}(II) demonstrates that broadband
low frequency squeezing was generated at sideband frequencies as low as 30~kHz.
A few extra noise peaks at 40 and 70 kHz  were most likely due to
backscattering of the pump field into the system.
%



In conclusion, we successfully  demonstrated     generation   of low-noise
broadband squeezing via self-rotation effect in  $^{87}$Rb at 795~nm,
independently confirming previous proof-of-principle experiments of Ries
\emph{et al.}~\cite{ries_experimental_2003}. The maximal measured squeezing
value is $0.87\pm0.02$~dB at  400~kHz, and the squeezing was detected in the
range of sideband frequencies from 30~kHz to 1.2~MHz. Up to our best knowledge,
this is the first demonstration of a sub-MHz quadrature-squeezed vacuum in
atomic systems. Further improvements are possible with optimization of the
system (\emph{i.e.} by optimizing the buffer gas pressure and composition, pump
field parameters, etc.). Supporting detail theoretical analysis is in progress.
Such low-cost method for generation of low-frequency non-classical fields near
atomic optical resonances may be useful and attractive for various quantum
memory tests and applications.


We thank Andrey Matsko and Seth Aubin for valuable contributions. This research
was partially supported by Jeffress Research grant J-847.


\bibliographystyle{osajnl}  

\begin{thebibliography}{10}
\newcommand{\enquote}[1]{``#1''}

\bibitem{lukin03rmp}
M.~D. Lukin, \enquote{Colloquium: Trapping and manipulating photon states in
  atomic ensembles,} Reviews of Modern Physics \textbf{75}, 457 (2003).

\bibitem{julsgaard04}
B.~Julsgaard, J.~Sherson, J.~I. Cirac, J.~Fiurasek, and E.~S. Polzik,
  \enquote{Experimental demonstration of quantum memory for light,} Nature
  \textbf{432}, 482--486 (2004).

\bibitem{chaneliere_storage_2005}
T.~Chaneliere, D.~N. Matsukevich, S.~D. Jenkins, S.-Y. Lan, T.~A.~B. Kennedy,
  and A.~Kuzmich, \enquote{Storage and retrieval of single photons transmitted
  between remote quantum memories,} Nature \textbf{438}, 833--836 (2005).

\bibitem{eisaman_electromagnetically_2005}
M.~D. Eisaman, A.~Andr\'{e}, F.~Massou, M.~Fleischhauer, A.~S. Zibrov, and
  M.~D. Lukin, \enquote{Electromagnetically induced transparency with tunable
  single-photon pulses,} Nature \textbf{438}, 837--841 (2005).

\bibitem{eisaman04}
M.~D. Eisaman, L.~Childress, A.~Andr\'{e}, F.~Massou, A.~S. Zibrov, and M.~D.
  Lukin, \enquote{Shaping quantum pulses of light via coherent atomic memory,}
  Physical Review Letters \textbf{93}, 233602 (2004).

\bibitem{matsukevich06}
D.~N. Matsukevich, T.~Chaneliere, S.~D. Jenkins, S.-Y. Lan, T.~A.~B. Kennedy,
  and A.~Kuzmich, \enquote{Deterministic single photons via conditional quantum
  evolution,} Physical Review Letters \textbf{97}, 013601--4 (2006).

\bibitem{chen06}
S.~Chen, Y.-A. Chen, T.~Strassel, Z.-S. Yuan, B.~Zhao, J.~Schmiedmayer, and
  J.-W. Pan, \enquote{Deterministic and storable single-photon source based on
  a quantum memory,} Physical Review Letters \textbf{97}, 173004--4 (2006).

\bibitem{kolchin_generation_2006}
P.~Kolchin, S.~Du, C.~Belthangady, G.~Y. Yin, and S.~E. Harris,
  \enquote{Generation of narrow-bandwidth paired photons: Use of a single
  driving laser,} Physical Review Letters \textbf{97}, 113602--4 (2006).

\bibitem{bachor_guide_2004}
H.-A. Bachor and T.~C. Ralph, \emph{A Guide to Experiments in Quantum Optics}
  (Wiley-VCH, 2004), 2nd ed.

\bibitem{hetet_squeezed_at_D1_Rb_2007}
G.~H{\'e}tet, O.~Gl{\"o}ckl, K.~A. Pilypas, C.~C. Harb, B.~C. Buchler, H.-A.
  Bachor, and P.~K. Lam, \enquote{Squeezed light for bandwidth-limited atom
  optics experiments at the rubidium D1 line,} Journal of Physics B \textbf{40}, 221--226 (2007).

\bibitem{chelkowski_10Hz_squeezing_2007}
S.~Chelkowski, H.~Vahlbruch, K.~Danzmann, and R.~Schnabel, \enquote{Coherent
  control of broadband vacuum squeezing,} Physical Review A \textbf{75}, 043814--9 (2007).

\bibitem{akamatsu_ultraslow_2007}
D.~Akamatsu, Y.~Yokoi, M.~Arikawa, S.~Nagatsuka, T.~Tanimura, A.~Furusawa, and
  M.~Kozuma, \enquote{Ultraslow propagation of squeezed vacuum pulses with
  electromagnetically induced transparency,} Physical Review Letters
  \textbf{99}, 153602--4 (2007).

\bibitem{appel_sq_quantum_memory_Rb_2007}
J.~Appel, E.~Figueroa, D.~Korystov, and A.~I. Lvovsky, \enquote{Quantum memory
  for squeezed light,} arXiv:0709.2258  (2007).

\bibitem{honda_sq_storage_in_Rb_2007}
K.~Honda, D.~Akamatsu, M.~Arikawa, Y.~Yokoi, K.~Akiba, S.~Nagatsuka,
  T.~Tanimura, A.~Furusawa, and M.~Kozuma, \enquote{Storage and retrieval of a
  squeezed vacuum,} arXiv:0709.1785  (2007).

\bibitem{lezama_numerical_2008}
A.~Lezama, P.~Valente, H.~Failache, M.~Martinelli, and P.~Nussenzveig,
  \enquote{Numerical investigation of the quantum fluctuations of optical
  fields transmitted through an atomic medium,} Physical Review A \textbf{77}, 013806--11 (2008).

\bibitem{matsko_vacuum_2002}
A.~B. Matsko, I.~Novikova, G.~R. Welch, D.~Budker, D.~F. Kimball, and S.~M.
  Rochester, \enquote{Vacuum squeezing in atomic media via self-rotation,}
  Physical Review A \textbf{66}, 043815 (2002).

\bibitem{ries_experimental_2003}
J.~Ries, B.~Brezger, and A.~I. Lvovsky, \enquote{Experimental vacuum squeezing
  in rubidium vapor via self-rotation,} Physical Review A \textbf{68}, 025801
  (2003).

\bibitem{hsu_effect_2006}
M.~T.~L. Hsu, G.~Hetet, A.~Peng, C.~C. Harb, H.-A. Bachor, M.~T. Johnsson,
  J.~J. Hope, P.~K. Lam, A.~Dantan, J.~Cviklinski, A.~Bramati, and M.~Pinard,
  \enquote{Effect of atomic noise on optical squeezing via polarization
  self-rotation in a thermal vapor cell,} Physical Review A  \textbf{73}, 023806--9 (2006).

\bibitem{novikova_ac-stark_2000}
I.~Novikova, A.~B. Matsko, V.~A. Sautenkov, V.~L. Velichansky, G.~R. Welch, and
  M.~O. Scully, \enquote{Ac-stark shifts in the nonlinear Faraday effect,} Opt.
  Lett. \textbf{25}, 1651--1653 (2000).

\bibitem{rochester_self-rotation_2001}
S.~M. Rochester, D.~S. Hsiung, D.~Budker, R.~Y. Chiao, D.~F. Kimball, and V.~V.
  Yashchuk, \enquote{Self-rotation of resonant elliptically polarized light in
  collision-free rubidium vapor,} Physical Review A \textbf{63}, 043814 (2001).

\bibitem{mckenzie_quantum_2005}
K.~McKenzie, E.~E. Mikhailov, K.~Goda, P.~K. Lam, N.~Grosse, M.~B. Gray,
  N.~Mavalvala, and D.~E. McClelland, \enquote{Quantum noise locking,} Journal
  of Optics B \textbf{7}, S421--S428 (2005).

\end{thebibliography}


\end{document}